\pdfoutput=1
\documentclass[12pt,a4paper]{article}

\usepackage{ifthen} % for conditional statements
\newboolean{pdflatex}
\setboolean{pdflatex}{true} % False for eps figures 

\newboolean{articletitles}
\setboolean{articletitles}{true} % False removes titles in references

\newboolean{uprightparticles}
\setboolean{uprightparticles}{false} %True for upright particle symbols

\def\paperauthors   {LHCb collaboration} 
\def\paperasciititle{Observation of a new charmed baryon decaying to Xic+ pi- pi+} 
\def\papertitle     {Observation of a new charmed baryon decaying to $\Xicp \pim \pip$} 
\def\paperkeywords  {{High Energy Physics}, {LHCb}} 
\def\papercopyright {\the\year\ CERN for the benefit of the LHCb collaboration} 
\def\paperlicence   {CC BY 4.0 licence}
\def\paperlicenceurl{https://creativecommons.org/licenses/by/4.0/}

\newif\ifEnableSectionTOCLinks
\EnableSectionTOCLinksfalse % deactivated

\usepackage[top=1in, bottom=1.25in, left=1in, right=1in]{geometry}

\usepackage{microtype}
\usepackage{lineno}  % for line numbering during review
\usepackage{xspace} % To avoid problems with missing or double spaces after
\usepackage{caption} %these three command get the figure and table captions automatically small

\usepackage{graphicx}  % to include figures (can also use other packages)
\usepackage{color}
\usepackage{colortbl}
\graphicspath{{./figs/}} % Make Latex search fig subdir for figures
\usepackage{amsmath} % Adds a large collection of math symbols
\usepackage{amssymb}
\usepackage{amsfonts}
\usepackage{upgreek} % Adds in support for greek letters in roman typeset

\newcommand*\patchAmsMathEnvironmentForLineno[1]{%
\expandafter\let\csname old#1\expandafter\endcsname\csname #1\endcsname
\expandafter\let\csname oldend#1\expandafter\endcsname\csname
end#1\endcsname
 \renewenvironment{#1}%
   {\linenomath\csname old#1\endcsname}%
   {\csname oldend#1\endcsname\endlinenomath}%
}
\newcommand*\patchBothAmsMathEnvironmentsForLineno[1]{%
  \patchAmsMathEnvironmentForLineno{#1}%
  \patchAmsMathEnvironmentForLineno{#1*}%
}
\AtBeginDocument{%
\patchBothAmsMathEnvironmentsForLineno{equation}%
\patchBothAmsMathEnvironmentsForLineno{align}%
\patchBothAmsMathEnvironmentsForLineno{flalign}%
\patchBothAmsMathEnvironmentsForLineno{alignat}%
\patchBothAmsMathEnvironmentsForLineno{gather}%
\patchBothAmsMathEnvironmentsForLineno{multline}%
\patchBothAmsMathEnvironmentsForLineno{eqnarray}%
}

\usepackage{hyperxmp}

\usepackage[pdftex,
            pdfauthor={\paperauthors},
            pdftitle={\paperasciititle},
            pdfkeywords={\paperkeywords},
            pdfcopyright={Copyright (C) \papercopyright},
            pdflicenseurl={\paperlicenceurl}]{hyperref}
\usepackage[bottom,flushmargin,hang,multiple]{footmisc}

\usepackage[all]{hypcap} % Internal hyperlinks to floats.

\usepackage{xspace} 
\usepackage{upgreek}

\def\lhcb   {\mbox{LHCb}\xspace}

\def\MagUp {\mbox{\em Mag\kern -0.05em Up}\xspace}

\ifthenelse{\boolean{uprightparticles}}%
{
 
 \def\Pgamma      {\ensuremath{\upgamma}\xspace}

 \def\Ppi         {\ensuremath{\uppi}\xspace}

 \def\PDelta      {\ensuremath{\Delta}\xspace}                 
 \def\PXi         {\ensuremath{\Xi}\xspace}                 
 \def\PLambda     {\ensuremath{\Lambda}\xspace}                 
 \def\PSigma      {\ensuremath{\Sigma}\xspace}                 
 \def\POmega      {\ensuremath{\Omega}\xspace}                 
 \def\PUpsilon    {\ensuremath{\Upsilon}\xspace}
 \let\oldPi\Pi
 \def\PPi         {\ensuremath{\oldPi}\xspace}

 \def\PB      {\ensuremath{\mathrm{B}}\xspace}                 
 \def\PD      {\ensuremath{\mathrm{D}}\xspace}                 

 \def\PK      {\ensuremath{\mathrm{K}}\xspace}                 
 \def\Pb      {\ensuremath{\mathrm{b}}\xspace}                 
 \def\Pc      {\ensuremath{\mathrm{c}}\xspace}                 
 \def\Pd      {\ensuremath{\mathrm{d}}\xspace}                 

 \def\Pp      {\ensuremath{\mathrm{p}}\xspace}                 

 \def\Ps      {\ensuremath{\mathrm{s}}\xspace}                 
                  
 \def\Pu      {\ensuremath{\mathrm{u}}\xspace}                 
 \def\thebaroffset{0.0em}
}
{
 
 \def\Pgamma      {\ensuremath{\gamma}\xspace}

 \def\Ppi         {\ensuremath{\pi}\xspace}

 \mathchardef\PDelta="7101
 \mathchardef\PXi="7104
 \mathchardef\PLambda="7103
 \mathchardef\PSigma="7106
 \mathchardef\POmega="710A
 \mathchardef\PUpsilon="7107
 \mathchardef\PPi="7105
 \def\PB      {\ensuremath{B}\xspace}                 
 \def\PD      {\ensuremath{D}\xspace}                 

 \def\PK      {\ensuremath{K}\xspace}                 
 \def\Pb      {\ensuremath{b}\xspace}                 
 \def\Pc      {\ensuremath{c}\xspace}                 
 \def\Pd      {\ensuremath{d}\xspace}                 

 \def\Pp      {\ensuremath{p}\xspace}                 

 \def\Ps      {\ensuremath{s}\xspace}                 
                  
 \def\Pu      {\ensuremath{u}\xspace}                 
 \def\thebaroffset{0.18em}
}
\newcommand{\offsetoverline}[2][\thebaroffset]{\kern #1\overline{\kern -#1 #2}}%

\makeatletter
\ifcase \@ptsize \relax% 10pt
  \newcommand{\miniscule}{\@setfontsize\miniscule{4}{5}}% \tiny: 5/6
\or% 11pt
  \newcommand{\miniscule}{\@setfontsize\miniscule{5}{6}}% \tiny: 6/7
\or% 12pt
  \newcommand{\miniscule}{\@setfontsize\miniscule{5}{6}}% \tiny: 6/7
\fi
\makeatother

\DeclareRobustCommand{\optbar}[1]{\shortstack{{\miniscule (\rule[.5ex]{1.25em}{.18mm})}
  \\ [-.7ex] $#1$}}

\def\g      {{\ensuremath{\Pgamma}}\xspace}

\def\uquark    {{\ensuremath{\Pu}}\xspace}

\def\dquark    {{\ensuremath{\Pd}}\xspace}

\def\squark    {{\ensuremath{\Ps}}\xspace}

\def\cquark    {{\ensuremath{\Pc}}\xspace}

\def\bquark    {{\ensuremath{\Pb}}\xspace}

\def\pion   {{\ensuremath{\Ppi}}\xspace}
\def\piz    {{\ensuremath{\pion^0}}\xspace}
\def\pip    {{\ensuremath{\pion^+}}\xspace}
\def\pim    {{\ensuremath{\pion^-}}\xspace}
\def\pipm   {{\ensuremath{\pion^\pm}}\xspace}

\def\kaon    {{\ensuremath{\PK}}\xspace}
\def\KorKbar {\kern \thebaroffset\optbar{\kern -\thebaroffset \PK}{}\xspace}

\def\Km      {{\ensuremath{\kaon^-}}\xspace}

\def\D       {{\ensuremath{\PD}}\xspace}

\def\DorDbar {\kern \thebaroffset\optbar{\kern -\thebaroffset \PD}\xspace}

\def\Dm      {{\ensuremath{\D^-}}\xspace}

\def\DpDm    {\ensuremath{\Dp {\kern -0.16em \Dm}}\xspace}

\def\B       {{\ensuremath{\PB}}\xspace}

\def\BorBbar {\kern \thebaroffset\optbar{\kern -\thebaroffset \PB}\xspace}

\def\Bd      {{\ensuremath{\B^0}}\xspace}

\def\BdorBdbar {\kern \thebaroffset\optbar{\kern -\thebaroffset \Bd}\xspace}

\def\Bs      {{\ensuremath{\B^0_\squark}}\xspace}

\def\BsorBsbar {\kern \thebaroffset\optbar{\kern -\thebaroffset \Bs}\xspace}

\def\Y#1S{\ensuremath{\PUpsilon{(#1S)}}\xspace}

\def\proton      {{\ensuremath{\Pp}}\xspace}

\def\Lz          {{\ensuremath{\PLambda}}\xspace}

\def\LorLbar     {\kern \thebaroffset\optbar{\kern -\thebaroffset \PLambda}\xspace}

\def\Xires       {{\ensuremath{\PXi}}\xspace}

\def\Lc          {{\ensuremath{\Lz^+_\cquark}}\xspace}

\def\Xic         {{\ensuremath{\Xires_\cquark}}\xspace}

\def\Xicp        {{\ensuremath{\Xires^+_\cquark}}\xspace}

\newcommand{\decay}[2]{\mbox{\ensuremath{#1\!\to #2}}\xspace} 

\def\to                 {\ensuremath{\rightarrow}\xspace}

\def\AT#1     {\ensuremath{A_{\mathrm{T}}^{#1}}\xspace}           % 2

\def\C#1      {\ensuremath{\mathcal{C}_{#1}}\xspace}                       % 9
\def\Cp#1     {\ensuremath{\mathcal{C}_{#1}^{'}}\xspace}                    % 7
\def\Ceff#1   {\ensuremath{\mathcal{C}_{#1}^{\mathrm{(eff)}}}\xspace}        % 9  
\def\Cpeff#1  {\ensuremath{\mathcal{C}_{#1}^{'\mathrm{(eff)}}}\xspace}       % 7
\def\Ope#1    {\ensuremath{\mathcal{O}_{#1}}\xspace}                       % 2
\def\Opep#1   {\ensuremath{\mathcal{O}_{#1}^{'}}\xspace}                    % 7

\newcommand{\aunit}[1]{\ensuremath{\text{\,#1}}}       
\newcommand{\tev}{\aunit{Te\kern -0.1em V}\xspace}
\newcommand{\gev}{\aunit{Ge\kern -0.1em V}\xspace}
\newcommand{\mev}{\aunit{Me\kern -0.1em V}\xspace}
\newcommand{\kev}{\aunit{ke\kern -0.1em V}\xspace}
\newcommand{\ev}{\aunit{e\kern -0.1em V}\xspace}
 
\newcommand{\mevc}{\ensuremath{\aunit{Me\kern -0.1em V\!/}c}\xspace}
\newcommand{\gevc}{\ensuremath{\aunit{Ge\kern -0.1em V\!/}c}\xspace}
\newcommand{\mevcc}{\ensuremath{\aunit{Me\kern -0.1em V\!/}c^2}\xspace}
\newcommand{\gevcc}{\ensuremath{\aunit{Ge\kern -0.1em V\!/}c^2}\xspace}
\def\fb   {\ensuremath{\aunit{fb}}\xspace}
\def\invfb   {\ensuremath{\fb^{-1}}\xspace}

\def\ps   {\ensuremath{\aunit{ps}}\xspace}

\def\gsim{{~\raise.15em\hbox{$>$}\kern-.85em
          \lower.35em\hbox{$\sim$}~}\xspace}
\def\lsim{{~\raise.15em\hbox{$<$}\kern-.85em
          \lower.35em\hbox{$\sim$}~}\xspace}

\def\sPlot{\mbox{\em sPlot}\xspace}

\def\pt         {\ensuremath{p_{\mathrm{T}}}\xspace}

\def\mrad{\aunit{mrad}\xspace}

\def\evtgen     {\mbox{\textsc{EvtGen}}\xspace}

\def\geant      {\mbox{\textsc{Geant4}}\xspace}

\def\pythia     {\mbox{\textsc{Pythia}}\xspace}

\def\tell1  {TELL1\xspace}
\def\ukl1   {UKL1\xspace}

\newcommand{\eg}{\mbox{\itshape e.g.}\xspace}
\newcommand{\ie}{\mbox{\itshape i.e.}\xspace}

\newcommand{\lhcborcid}[1]{\href{https://orcid.org/#1}{\hspace*{0.1em}\raisebox{-0.45ex}{\includegraphics[width=1em]{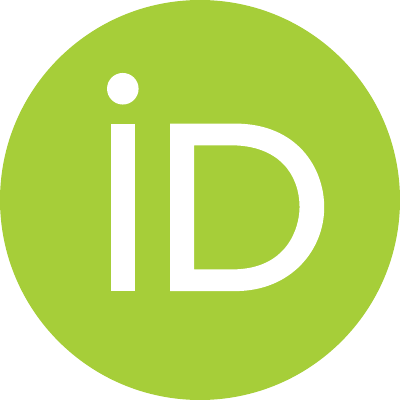}}}}

\hypersetup{
  colorlinks   = true, %Colours links instead of ugly boxes
  urlcolor     = blue, %Colour for external hyperlinks
  linkcolor    = blue, %Colour of internal links
  citecolor    = red   %Colour of citations
}

\ifEnableSectionTOCLinks
    \usepackage[explicit]{titlesec} % to change headings
    
    \let\oldcontentsline\contentsline
    \renewcommand

    \titleformat{\section}{\normalfont\Large\bf}{\hyperlink{tocsection.\thesection}{{\thesection} \parbox[t]{\dimexpr\textwidth-1pc}{#1}}}{1pc}{}

    \titleformat{\subsection}{\normalfont\bf}{\hyperlink{tocsubsection.\thesubsection}{{\thesubsection} \parbox[t]{\dimexpr\textwidth-1pc}{#1}}}{1pc}{}

    \titleformat{name=\section,numberless}[display]{}{}{0pt}{\normalfont\Huge\bfseries #1}
\fi

\usepackage{cite} % Allows for ranges in citations
\usepackage{mciteplus}
\newcommand{\Xicstst}[1]{\ensuremath{\Xic(#1)}\xspace} 
\newcommand{\Xicststp}[1]{\ensuremath{\Xic(#1)^+}\xspace} 
\newcommand{\Xicststz}[1]{\ensuremath{\Xic(#1)^0}\xspace} 

\def\Xicpst      {{\ensuremath{\Xires^{**+}_\cquark}}\xspace}

\def\Xicprime        {{\ensuremath{\Xires^\prime_\cquark}}\xspace}

\usepackage{multirow}
\def\XicOneDelta    {\ensuremath{  69.80 \pm    0.03 \pm   0.03           \mev}}
\def\XicOneMass     {\ensuremath{2816.65 \pm    0.03 \pm   0.03 \pm 0.23  \mev}}
\def\XicOneWidth    {\ensuremath{   2.07 \pm    0.08 \pm   0.12           \mev}}

\def\XicTwoDelta    {\ensuremath{ 175.95 \pm    0.28 \pm   0.54           \mev}}
\def\XicTwoMass     {\ensuremath{2922.8  \pm    0.3  \pm   0.5 \pm 0.2  \mev}}
\def\XicTwoWidth    {\ensuremath{   5.3  \pm    0.9  \pm   1.4            \mev}}

\def\XicThreeDelta  {\ensuremath{ 221.77 \pm    0.46 \pm   0.48           \mev}}
\def\XicThreeMass   {\ensuremath{2968.6  \pm    0.5  \pm   0.5 \pm 0.2  \mev}}
\def\XicThreeWidth  {\ensuremath{  31.7  \pm    1.7  \pm   1.9            \mev}}

\def\XicFourDelta   {\ensuremath{ 329.92 \pm    0.72 \pm   1.27           \mev}}
\def\XicFourMass    {\ensuremath{3076.8  \pm    0.7  \pm   1.3 \pm 0.2  \mev}}
\def\XicFourWidth   {\ensuremath{   6.8  \pm    2.3  \pm   0.9            \mev}}

\begin{document}

%%%%%%%%%%%%%%%%%%%%%%%%%
%%%%% Title     %%%%%%%%%
%%%%%%%%%%%%%%%%%%%%%%%%%

\renewcommand{\thefootnote}{\fnsymbol{footnote}}
\setcounter{footnote}{1}

% ===============================================================================
% Purpose: LHCb-PAPER journal paper title page template
% Author: 
% Created on: 2010-09-25
% ===============================================================================

%%%%%%%%%%%%%%%%%%%%%%%%%
%%%%%  TITLE PAGE  %%%%%%
%%%%%%%%%%%%%%%%%%%%%%%%%
\begin{titlepage}
\pagenumbering{roman}

% Header ---------------------------------------------------
\vspace*{-1.5cm}
\centerline{\large EUROPEAN ORGANIZATION FOR NUCLEAR RESEARCH (CERN)}
\vspace*{1.5cm}
\noindent
\begin{tabular*}{\linewidth}{lc@{\extracolsep{\fill}}r@{\extracolsep{0pt}}}
\vspace*{-1.5cm}\mbox{\!\!\!\includegraphics[width=.14\textwidth]{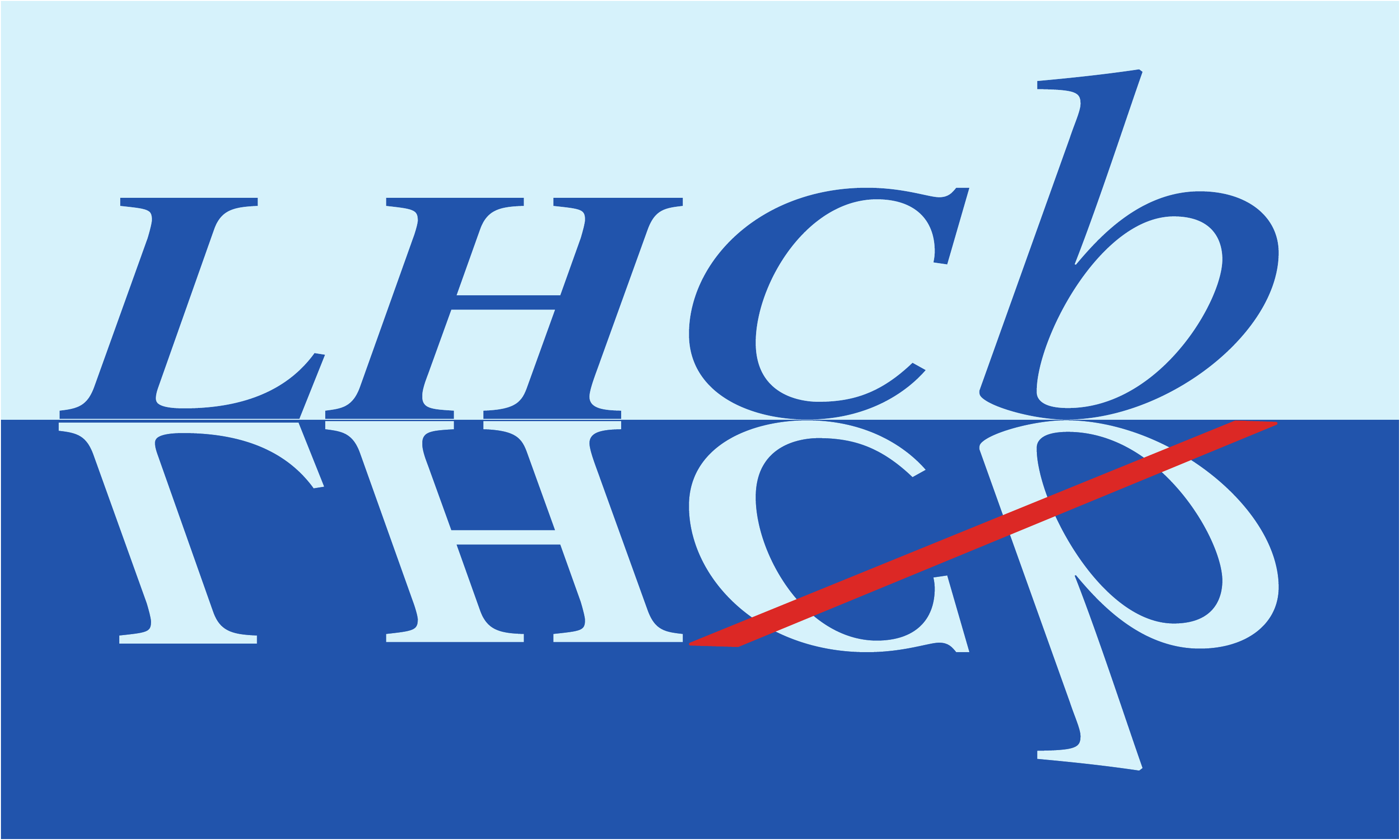}} & &
\\
 & & CERN-EP-2025-019 \\  % ID 
 & & LHCb-PAPER-2024-055 \\  % ID 
 & & October 22, 2025 \\ % Date - Can also hardwire e.g.: 23 March 2010
 & & \\
% not in paper \hline
\end{tabular*}

\vspace*{2.5cm}

% Title --------------------------------------------------
{\normalfont\bfseries\boldmath\huge
\begin{center}
% DO NOT EDIT HERE. Instead edit macro in main.tex to keep metadata correct
  \papertitle 
\end{center}
}

\vspace*{1.0cm}

% Authors -------------------------------------------------
\begin{center}
%In the footnote, replace 'paper' by 'Letter' in case of submission to PRL or PLB 
% Edit macro in main.tex to keep metadata correct
\paperauthors\footnote{Authors are listed at the end of this paper.}
\end{center}

\vspace{\fill}

% Abstract -----------------------------------------------
\begin{abstract}
  \noindent
The $\Xicp\pim\pip$ spectrum is investigated using proton-proton collisions at a center-of-mass energy of $13\tev$, corresponding to an integrated luminosity of $5.4\invfb$, collected by the $\lhcb$ experiment during $2016$--$2018$.
Four states are observed with high significance, and their masses and widths are measured to be
\begin{align*}
    m[\Xic(2815)^{+}]        &=  \XicOneMass,\\
    \Gamma[\Xic(2815)^{+}]   &=  \XicOneWidth,\\[5pt]
    m[\Xic(2923)^{+}]        &=    \XicTwoMass,\\
    \Gamma[\Xic(2923)^{+}]   &=    \XicTwoWidth,\\[5pt]
    m[\Xic(2970)^{+}]        &=    \XicThreeMass,\\
    \Gamma[\Xic(2970)^{+}]   &=    \XicThreeWidth,\\[5pt]
    m[\Xic(3080)^{+}]        &=    \XicFourMass,\\
    \Gamma[\Xic(3080)^{+}]   &=    \XicFourWidth,
\end{align*}
where the uncertainties are statistical, systematic, and due to the limited precision on the \Xicp mass, respectively. The $\Xic(2923)^{+}$ baryon is observed for the first time, and is consistent with being the isospin partner of the previously observed $\Xic(2923)^{0}$ state. Most of the measured parameters are more precise than existing world averages.

\end{abstract}

\vspace*{1.0cm}

\begin{center}
Published in Phys. Rev. Lett. 135 (2025) 161901
\end{center}

\vspace{\fill}

{\footnotesize 
% Edit macro in main.tex to keep metadata correct
\centerline{\copyright~\papercopyright. \href{\paperlicenceurl}{\paperlicence}.}}
\vspace*{2mm}

\end{titlepage}

%%%%%%%%%%%%%%%%%%%%%%%%%%%%%%%%
%%%%%  EOD OF TITLE PAGE  %%%%%%
%%%%%%%%%%%%%%%%%%%%%%%%%%%%%%%%

%  empty page follows the title page ----
\newpage
\setcounter{page}{2}
\mbox{~}
%\newpage
%
%% Author List ----------------------------
%%  You need to get a new author list!
%\input{LHCb_authorlist.tex}
%
%The author list for journal publications is provided by the Membership Committee shortly after 'approval to go to paper' has been given.
%%It will be made available on the page
%%\verb!http://www.physik.uzh.ch/~strauman/forMemCo/LHCb-PAPER-XXXX-XXX/! .
%It will be sent to you by email shortly after a paper number has beens assigned.
%The author list should be included already at first circulation, 
%to allow new members of the collaboration to verify whether they have been included correctly.
%Occasionally a misspelled name is corrected or associated institutions become full members.
%In that case, a new author list will be sent to you.
%In case line numbering doesn't work well after including the authorlist, try moving the \verb!\bigskip! after the last author to a separate line.
%
%
%The authorship for Conference Reports should be ``The LHCb
%  collaboration'', with a footnote giving the name(s) of the contact
%  author(s), but without the full list of collaboration names.

\renewcommand{\thefootnote}{\arabic{footnote}}
\setcounter{footnote}{0}

\cleardoublepage

%%%%%%%%%%%%%%%%%%%%%%%%%
%%%%% Main text %%%%%%%%%
%%%%%%%%%%%%%%%%%%%%%%%%%

\pagestyle{plain} % restore page numbers for the main text
\setcounter{page}{1}
\pagenumbering{arabic}

%% Comment before a final submission.
%\linenumbers

Singly charmed baryons are bound states of a charm quark (\cquark) and two light quarks (\uquark, \dquark or \squark).
Due to the large mass difference between the charm and the lighter quarks, these baryons provide an insight into the spectrum of hadronic states using symmetries described by the Heavy
Quark Effective Theory~\cite{Grozin:1992yq,Mannel:1996rg}. Numerous theoretical predictions of the properties of heavy baryons, containing either a charm or a beauty quark, have been made in recent years~\cite{Migura:2006ep,Ebert:2007nw, Roberts:2007ni, Garcilazo:2007eh,   Valcarce:2008dr, Ebert:2011kk,  Vijande:2012mk, Yoshida:2015tia, Chen:2015kpa, Shah:2016nxi, Chen:2016phw, Oudichhya:2021kop, Jakhad:2023mni, Garcia-Tecocoatzi:2022zrf, Ortiz-Pacheco:2023kjn, Li:2023gbo, Weng:2024roa, Garcia-Tecocoatzi:2023btk,
Yang:2021lce,
Yang:2020zjl,
Yu:2018yxl,
Nieves:2019jhp,
Zhu:2020jke,
Wang:2017vtv},  
 including some based on Lattice QCD calculations~\cite{Padmanath:2013bla, Padmanath:2017lng, Chen:2017kxr, Bahtiyar:2020uuj}.
There are fifteen singly charmed baryons with orbital angular momentum $L=0$, all of which have now been observed~\cite{PDG2024}. 
The number increases drastically when considering the orbital and radial excitations. Modeling such states as a heavy quark interacting with a light diquark can help restrict the number of possible physical states~\cite{Garcia-Tecocoatzi:2022zrf, Garcia-Tecocoatzi:2023btk}.  However, despite the progress made in recent years, this number remains much higher than that observed experimentally~\cite{LHCb-PAPER-2016-061, LHCb-PAPER-2017-002, Yelton:2017qxg, Li:2017uvv, LHCb-PAPER-2020-004, LHCb-PAPER-2022-043}. This implies that many other states remain undiscovered or that some configurations are forbidden by selection rules. The latter would have a broad impact in hadron spectroscopy given that the diquark is also used as a building block of exotic states, including tetraquarks and pentaquarks~\cite{Jaffe:1976ig, Jaffe:1976ih, Maiani:2004vq}.

The \Xic baryons, states with quark content \cquark\squark\uquark or \cquark\squark\dquark, are described by a wave function which is antisymmetric (\eg the \Xic ground state) or symmetric (\eg~\Xicprime or \Xicstst{2645} excitations) under interchange of the light quark flavors. More than one hundred excited \Xic baryons are expected in the mass region $2.8$--$3.8\gev$~\cite{Padmanath:2013bla} and the mass splitting between the states can be so small, relative to their natural widths, that they are difficult to resolve experimentally.\footnote{Natural units with $\hbar=c=1$ are used throughout.}
This might be the case for the \Xicstst{2970} state whose natural width was measured to be about 30\mev in the \Xicprime \Ppi and \mbox{\Xic \pip \pim}   
spectra~\cite{Belle:2016lhy} while only 15\mev in the \Lc \Km \pip\cite{Belle:2013htj} and \Lc \Km~\cite{LHCb-PAPER-2020-004} systems. More precise measurements are needed to determine if two different \Xic baryons have been mistakenly identified as a single state due to the proximity of their masses.

This Letter presents a search for excited \Xicp baryons, hereafter collectively termed \Xicpst, in the \Xicp\pim\pip spectrum. This final state was studied by the Belle collaboration, which reported the observation of the \Xicststp{2815} and \Xicststp{2970} baryons~\cite{Belle:2016lhy}. 
 The spin-parity of the latter was also measured to be $J^P = \frac{1}{2}^+$~\cite{Belle:2020tom}. A further structure peaking at around 2.92\gev was noted, although it was not statistically significant. 
 
Data used in this analysis were collected by the \lhcb experiment during 2016--2018 from proton-proton ($pp$) collisions at a center-of-mass energy of 13\tev, corresponding to an integrated luminosity of 5.4\invfb. Excited \Xicp states are reconstructed through their intermediate decay to $\Xicststz{2645}\pip$, with $\decay{\Xicststz{2645}}{ \Xicp\pim}$ and $\decay{\Xicp}{ \proton\Km\pip}$.\footnote{The inclusion of charge-conjugate processes is implied throughout.} 

%%% Detector, trigger, simulation

The \lhcb detector~\cite{LHCb-DP-2008-001,LHCb-DP-2014-002} is a single-arm forward spectrometer covering the pseudorapidity range $2 < \eta < 5$, designed for
the study of particles containing \bquark or \cquark quarks. The detector elements particularly relevant to this analysis are: a silicon-strip vertex detector surrounding the $pp$ interaction region that allows \cquark\ and \bquark\ hadrons to be identified from their characteristically long flight distance~\cite{LHCb-DP-2014-001}; a large-area silicon-strip detector located upstream of a dipole magnet with a bending power of about $4{\mathrm{\,T\,m}}$, and three stations of silicon-strip detectors and straw drift tubes~\cite{LHCb-DP-2017-001} placed downstream of the magnet; and two ring-imaging Cherenkov detectors that provide particle identification (PID) information to discriminate between different species of charged hadrons~\cite{LHCb-DP-2012-003}.
The online event selection is performed by a trigger~\cite{LHCb-DP-2012-004} consisting of a hardware stage based on information from the calorimeter and muon systems, followed by a two-level software stage which applies a full event reconstruction.

Simulation is used in this analysis to optimize the event selection and to model the experimental mass resolution.
Proton-proton collisions are generated using \pythia~\cite{Sjostrand:2007gs,*Sjostrand:2006za} with a specific \lhcb configuration~\cite{LHCb-PROC-2010-056}. Decays of unstable particles are described by \evtgen~\cite{Lange:2001uf}.
The interaction of the generated particles with the detector, and its response, are implemented using the \geant toolkit~\cite{Allison:2006ve, *Agostinelli:2002hh} as described in  Ref.~\cite{LHCb-PROC-2011-006}. 

In order to maximize the signal efficiency, no explicit requirements are imposed at the hardware trigger stage. At the software trigger stage a dedicated algorithm reconstructs $\decay{\Xicp}{ \proton\Km\pip}$ candidates from the combination of three reconstructed tracks which must fulfill loose kinematic and PID requirements~\cite{Papanestis:2017zcj} and be consistent with originating from a common vertex with significant displacement from any primary interaction vertex (PV) in the event. 
The \Xicp candidate must have a measured lifetime with respect to the associated PV exceeding 0.15\ps, and a  mass in the range \mbox{$2432.5 < m(\proton\Km\pip) < 2502.5$\mev}.

%%% Event selection

The \Xicp candidates selected by the trigger are subsequently combined with a pair of oppositely charged particles, fulfilling loose pion PID requirements and consistent with originating from the \Xicp associated PV, to form $\decay{\Xicpst}{ \Xicststz{2645}} (\to \Xicp \pim)\pip$ candidates. The \Xicp\pim\pip system must form a good-quality vertex and satisfy the reconstructed mass conditions $m(\Xicp\pim\pip) < 3300\mev$ and [$m(\Xicp\pim) - m(\Xicp) - m^{\text{PDG}}_{\pim}] < 150\mev$, where the second condition selects the \Xicststz{2645} resonance region and $m^{\text{PDG}}_{\pim}$ is the known pion mass~\cite{PDG2024}.

Three selection stages are performed to sequentially optimize the signal for \Xicp, \Xicststz{2645}, and \Xicpst states. More information on each stage is included in the End Matter. In the first (\Xicp) selection stage, a series of loose criteria are initially imposed to reduce the dominant combinatorial background by 85\% while retaining 83\% of the signal. These include a requirement on the alignment of the \Xicp momentum with the direction determined by its production and decay vertices, PID requirements on each final-state particle, and loose requirements on topological variables related to the \Xicp flight distance and its consistency with originating from a PV.

\begin{figure}
    \centering
    \includegraphics[width=0.8\linewidth]{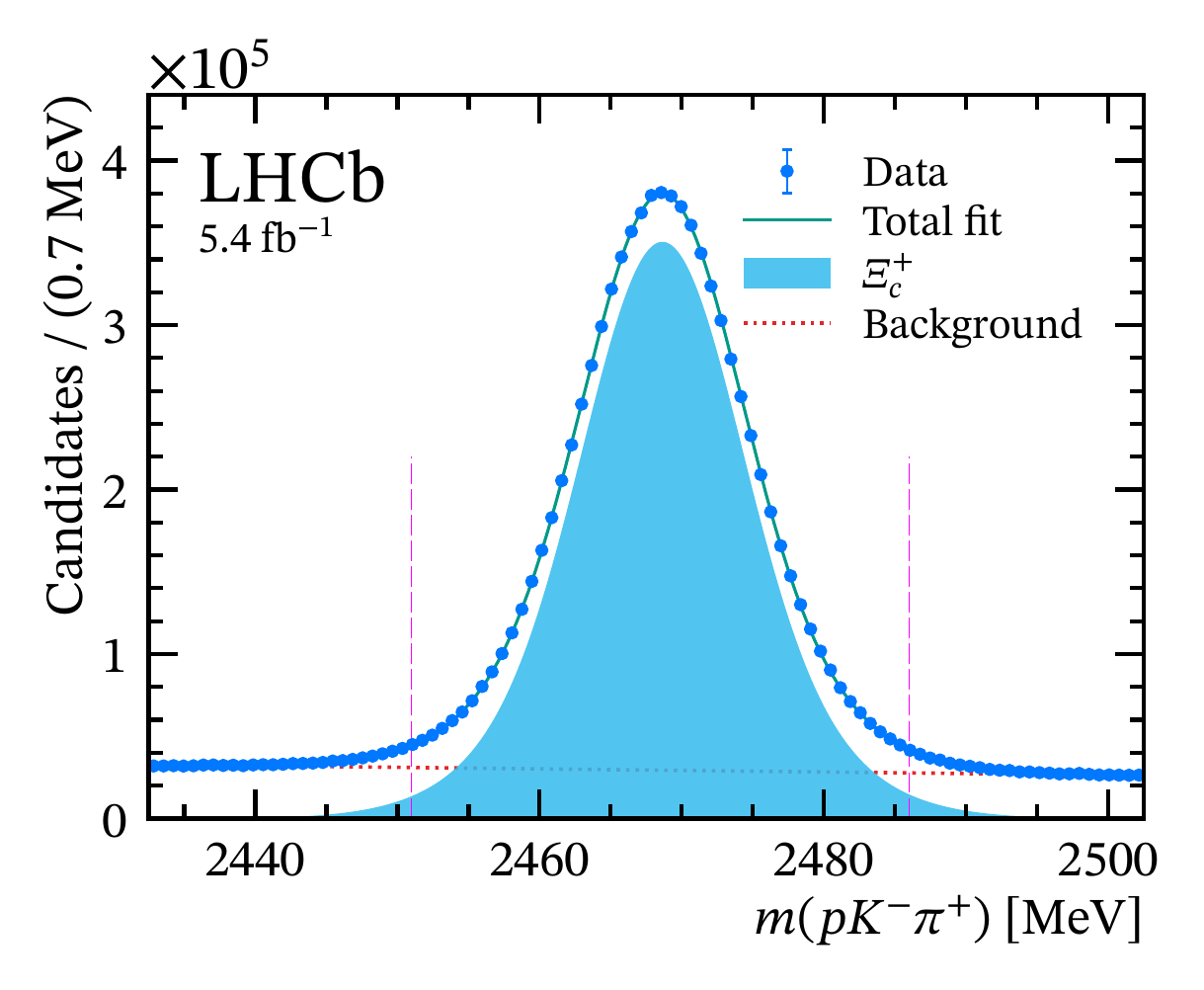}
    \caption{Distribution of $m(pK^-\pi^+)$ for candidates passing \Xicp selection requirements, together with the fit model. The dashed vertical lines delimit the mass window imposed for subsequent selection stages.}
    \label{fig:xicp}
\end{figure}

To further suppress background in the \Xicp sample, a multivariate algorithm (MVA) is trained with signal and background samples extracted using the \sPlot technique~\cite{Pivk:2004ty}. The \Xicp mass distribution is fitted to obtain weights for signal and background, based on a 5\% random sampling of the full data sample. In this fit the signal is described by a double Gaussian function, \ie the sum of two Gaussian functions constrained to have a common mean, and the background by a first-order polynomial function. A cross-validation method is used to ensure the classifier is not applied to the data on which it is trained. 
A feed-forward neural network~\cite{tensorflow2015-whitepaper, KERAS} is configured with a loss function that properly accounts for the weights in the training samples~\cite{Borisyak:2019vbz}. A total of 29 input variables are included, mainly topological quantities and PID variables for the final-state particles. Following the training, a requirement is imposed on the classifier output to maximize the signal significance $S/\sqrt{S+B}$, where $S$ ($B$) is the signal (background) yield. 
This selection retains 84\% (16\%) of the signal (background). The resulting $m(p\Km\pip)$ distribution, after imposing all \Xicp selection criteria, is shown in Fig.~\ref{fig:xicp} for the full data sample. Also shown is an indicative fit using the same model as described above. For subsequent analysis a tighter mass window of $2451.3 < m(p\Km\pip) < 2486.0$\mev is imposed, containing 98\% of the signal. Around $7.7 \times 10^6$ \Xicp signal candidates are selected, with a purity of 84\%.

In the second selection stage, a separate classifier is trained to optimize the signal significance of the \Xicststz{2645} state. The signal and background training samples are again derived from cross-validated data using the \sPlot technique~\cite{Pivk:2004ty}, following a fit to the quantity $m_{\text{corr}}[\Xicststz{2645}] \equiv m(\Xicp\pim) - m(\Xicp) + m^{\text{PDG}}_{\Xicp}$, where the final entry corresponds to the known \Xicp mass~\cite{PDG2024}. Using this mass difference leads to improved resolution and, hence, higher signal purity. In the fit, the \Xicststz{2645} signal peak is described by the convolution of a relativistic Breit--Wigner (RBW) function with a double Gaussian resolution function. Simulation is used to fix the parameters of the resolution function, with both widths free to vary in the fit through a common scale factor to account for possible mismodeling. The dominant background is combinatorial, either from genuine $\Xicp$ baryons combined with an unrelated track, or fake \Xicp candidates. These components are described empirically by a function $f(x) = a (x - m_0)^k + b (x - m_0)$ with $a$, $b$, $k$, and $m_0$ as free parameters. Two additional peaking backgrounds are observed in the sample. The first occurs close to the signal peak, originating from \mbox{$\decay{\Xicststz{2815}}{ \Xicststp{2645}} (\to \Xicp \piz) \pim$} decays, where the neutral pion is not reconstructed. The second occurs in the mass region around 2680\mev, due to $\decay{\Xicststz{2790}}{ \Xicprime^{+}} ( \to \Xicp \g) \pim$ decays with the photon not reconstructed. Both of these partially reconstructed backgrounds are described using templates derived from a fast simulation software~\cite{Cowan:2016tnm}. Their mass positions are allowed to shift through Gaussian constraints centered on zero and with widths 0.3\mev and 0.5\mev for the \Xicststz{2815} and \Xicststz{2790} decays, respectively, based on the experimental mass uncertainties of these states~\cite{PDG2024}.

\begin{figure}
    \centering
    \includegraphics[width=0.8\linewidth]{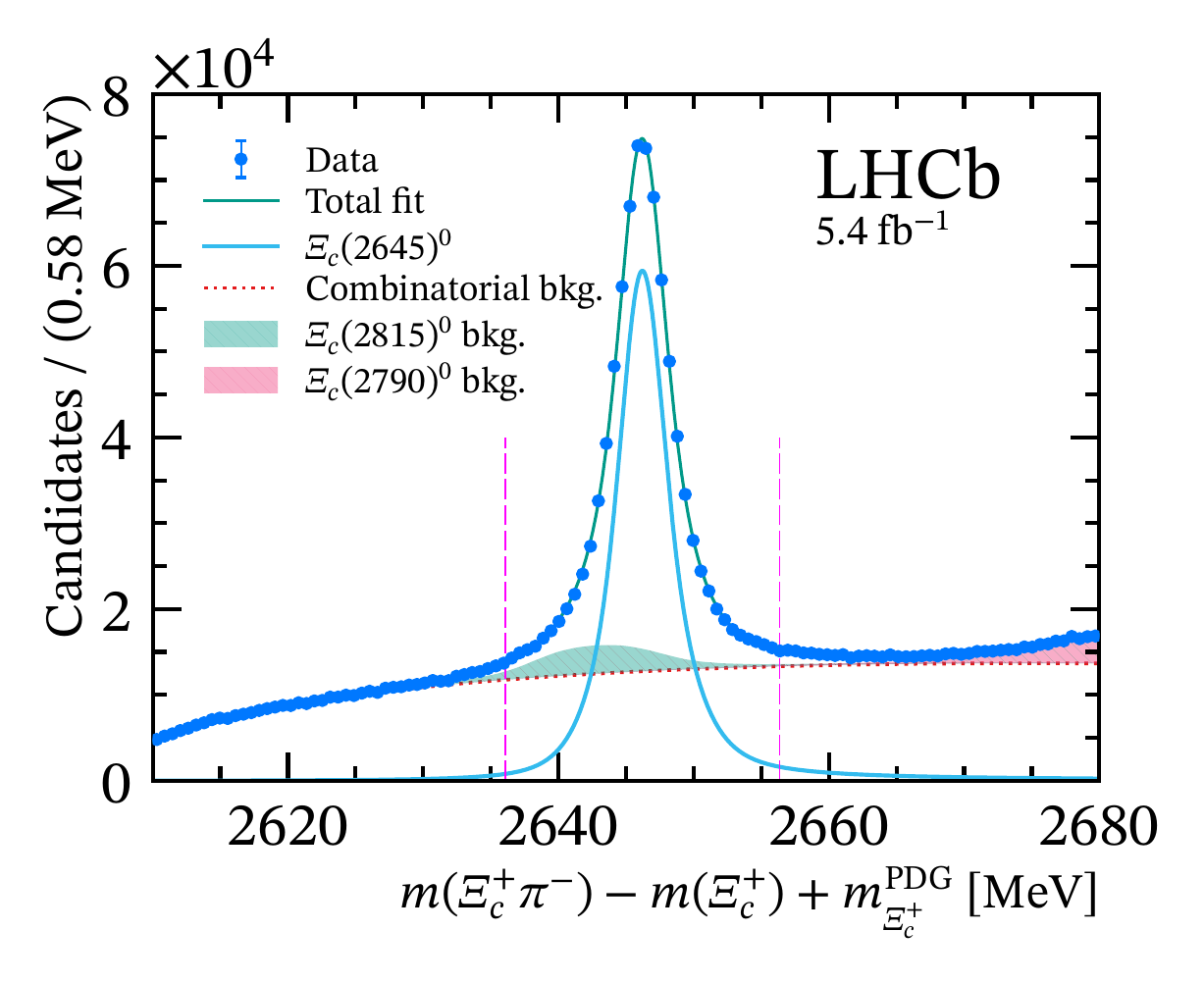}
    \caption{Distribution of $m_{\text{corr}}(\Xicststz{2645})$ for candidates passing the second selection stage. The fit model is also shown for illustrative purposes. The dashed vertical lines enclose the mass region selected for subsequent analysis.}
    \label{fig:xic2645}
\end{figure}

The weighted signal and background samples obtained from the \sPlot fit are used to train a gradient Boosted Decision Tree (BDT)~\cite{Breiman, AdaBoost,Scikit-learn-paper} with 14 input variables, one of which is the neural network output from the first selection stage. The other variables include PID and kinematic quantities for the pion, and topological quantities characterizing the \Xicp\pim vertex quality. A requirement is imposed on the BDT output to optimize the signal significance, which gives a signal (background) retention of 95\% (74\%). A mass requirement $2635 < m_{\text{corr}}[\Xicststz{2645}] < 2656\mev$ is then applied, containing 98\% of the signal candidates. In total, around $56.5\times 10^4$ \Xicststz{2645} signal candidates are selected, with a purity of 52\%. The mass distribution after all criteria have been imposed is shown in Fig.~\ref{fig:xic2645} along with the results of an illustrative fit with the same model as described above. 

The final selection stage uses a third MVA to optimize possible \Xicpst signal peaks in an unbiased way. Simulation is used as a proxy for the signal to train the classifier, with four samples generated with \Xicpst mass assignments 2815, 2923, 2970, and 3055\mev. The input variable distributions and classifier performance are similar for the different mass values, so these four samples are combined in the training. The background proxy sample is taken from data, combining the wrong-sign final states \Xicp\pip\pip and \Xicp\pim\pim, which is found to provide a good description of the background entering the signal sample. A gradient BDT~\cite{catboost} is trained with 15 input variables covering kinematic, topological, and PID quantities. Following training, a requirement on the BDT output is imposed to maximize the quantity $\varepsilon / (5 /2 + \sqrt{B})$~\cite{Punzi:2003bu} where $\varepsilon$ is the signal efficiency taken from simulation and $B$ is the background yield under the signal peak of interest, taken from the wrong-sign sample. The optimal requirement is calculated separately for the four mass assignments used in simulation, and exhibits a weak mass dependence. The chosen requirement is based on the \Xicststp{2923} hypothesis, and retains 34\% of signal while rejecting 99.3\% of background in the proxy samples.

A final requirement is imposed to reject cases where a single particle is reconstructed as two tracks, with both included in the final state. Such candidates are removed by requiring the two-dimensional angular separation between all pairs of final-state particles to exceed 0.08\mrad. Following the application of all selection criteria, 84\% of events include only one selected \Xicpst candidate. For events with multiple candidates, all are accepted, with an alternative treatment considered later as a cross-check.

%%% Mass fit

\begin{figure}
    \centering
    \includegraphics[width=\linewidth]{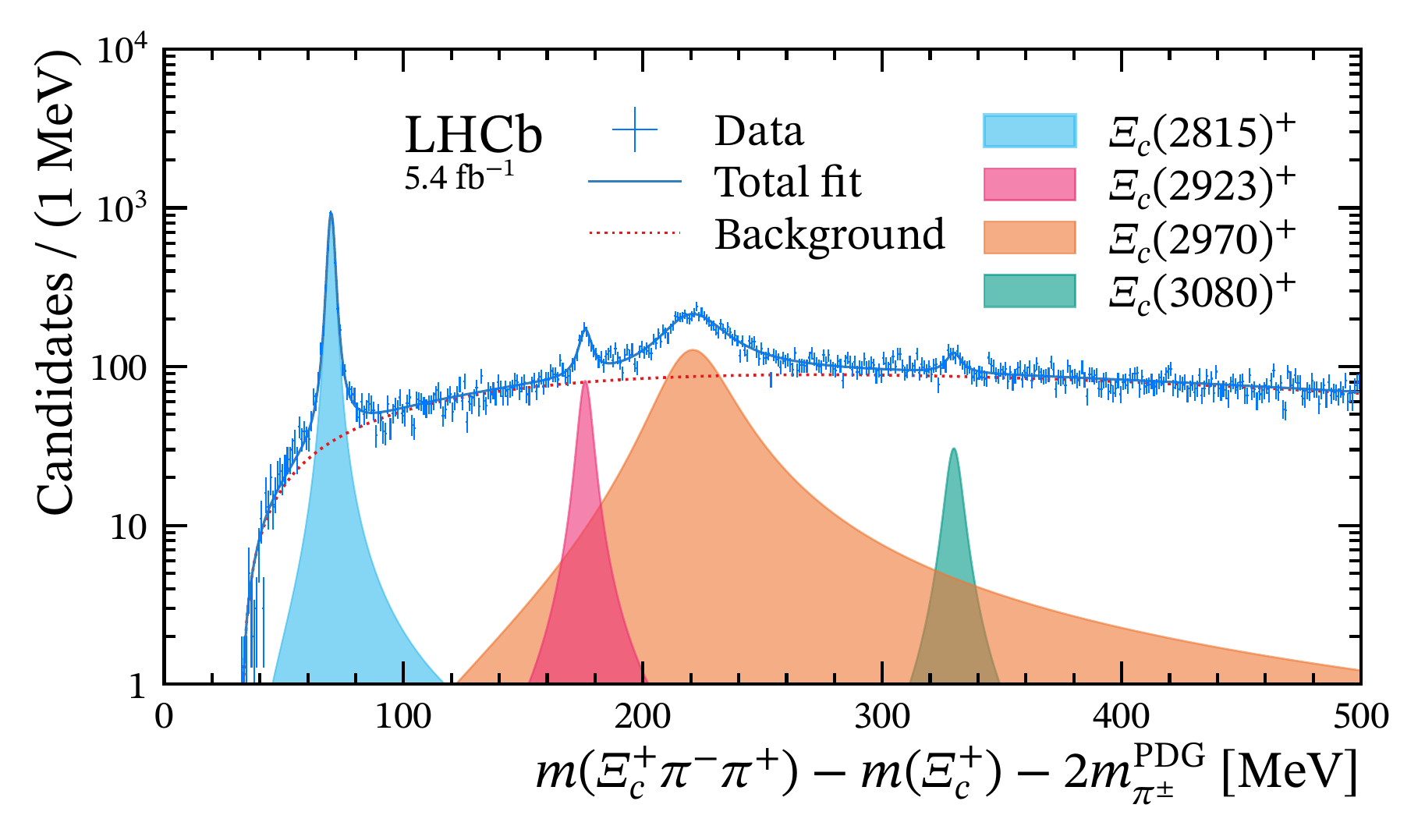}
    \caption{Distribution of $\Delta M$ for candidates passing all selection requirements, shown with the fit model. The individual signal components for the four observed states are drawn separately as filled curves, and the background as a dotted line.}
    \label{fig:DM}
\end{figure}

The mass distribution for candidates passing all selection requirements is shown in Fig.~\ref{fig:DM}, expressed in terms of the mass difference \mbox{$\Delta M \equiv m(\Xicp\pim\pip) - m(\Xicp) - 2\, m^{\text{PDG}}_{\pipm}$}. Four distinct peaks are observed above a smooth background rising from the kinematic threshold. These peaks are labeled as \Xicststp{2815}, \Xicststp{2923}, \Xicststp{2970}, and \Xicststp{3080} based on their proximity to existing states~\cite{PDG2024}, or in the case of \Xicststp{2923} to the corresponding neutral state observed in decays to $\Lc\Km$~\cite{LHCb-PAPER-2020-004}. 

An extended unbinned maximum-likelihood fit is performed on the $\Delta M$ distribution to determine the masses and widths of these states, and their statistical significances. In this fit, the background is described by an empirical function \mbox{$f(x) = x^k \exp(-c x)$}, where $x \equiv (\Delta M - m_0)$, and $k$, $c$, and $m_0$ are free parameters. This model choice is motivated by studies of the wrong-sign samples, giving a compromise between a good fit quality and a small number of free parameters.
Each signal peak is described by the convolution of a relativistic Breit--Wigner function with a resolution function obtained from simulation.
Different assumptions are tested for the Blatt--Weisskopf barrier factors \cite{Blatt:1952ije} and orbital angular momentum of the states, with the results used to assign systematic uncertainties as described later. 
The resolution function is determined using the four simulated \Xicpst samples, which correspond quite closely in mass to the states observed in data. 
The $\Delta M$ resolution is determined from simulation, and is described by a double Gaussian function with a mass-dependent scale factor applied to both widths. This resolution varies from 1\mev for the \Xicststp{2815} state to 2.5\mev for the \Xicststp{3080} state. 

The free parameters of the $\Delta M$ fit are the yields ($N$), peak positions, and widths ($\Gamma$) of each of the four signal peaks, the three background shape parameters, and the background yield. The absolute masses ($m$) are then determined from the peak positions in $\Delta M$ by adding the known \Xicp and \pipm masses~\cite{PDG2024}. The fit model provides a good description of the data, with a $\chi^2$ of 478 for 458 degrees of freedom. The parameters for the observed states are measured to be
\begin{align*}
    N[\Xic(2815)^{+}]        &=     4072 \pm    77, \\
    m[\Xic(2815)^{+}]        &=     \XicOneMass,    \\
    \Gamma[\Xic(2815)^{+}]   &=     \XicOneWidth,   \\[5pt]
    N[\Xic(2923)^{+}]        &=     738 \pm    76,  \\
    m[\Xic(2923)^{+}]        &=     \XicTwoMass,    \\
    \Gamma[\Xic(2923)^{+}]   &=     \XicTwoWidth,   \\[5pt]
    N[\Xic(2970)^{+}]        &=     6105 \pm   255, \\
    m[\Xic(2970)^{+}]        &=     \XicThreeMass,  \\
    \Gamma[\Xic(2970)^{+}]   &=     \XicThreeWidth, \\[5pt]
    N[\Xic(3080)^{+}]        &=     344 \pm    77,  \\
    m[\Xic(3080)^{+}]        &=     \XicFourMass,   \\
    \Gamma[\Xic(3080)^{+}]   &=     \XicFourWidth,
\end{align*}
where the first uncertainty is statistical, the second is due to experimental systematic effects discussed below, and the third uncertainty on the masses is from the limited knowledge of the \Xicp mass~\cite{PDG2024}.   

The statistical significance for the two lowest-yield peaks is determined from Wilks' theorem~\cite{Wilks:1938dza}. The \Xicststp{2923} state is observed with over 10$\sigma$ significance, representing the discovery of this state. The \Xicststp{3080} state is observed with 5.4$\sigma$ significance. To account for possible systematic effects from the choice of signal and background model, the calculation is repeated for the alternative model choices described below, and the lowest significance is quoted. 

%%% Systematic uncertainties and cross-checks

Various sources of systematic uncertainty are considered for the mass and width measurements, as summarized in Table~\ref{tab:syst}. Alternative descriptions of the signal peaks are tested by varying the Blatt--Weisskopf radius used in the RBW function from its baseline value of $3.1\gev^{-1}$ to $2.0\gev^{-1}$ and $4.0\gev^{-1}$. In addition, the assumed orbital angular momentum of the decay products, $L$, is varied independently for all four observed states to cover any physically allowed values between $L=0$ and $L=2$. The standard deviation of each signal parameter across the full ensemble of alternative signal models is taken as a systematic uncertainty (labeled \emph{RBW} in Table~\ref{tab:syst}). 
In the baseline fit no interference is included between the observed states. The \Xicststp{2970} resonance has a significant overlap with both the \Xicststp{2923} and \Xicststp{3080} peaks, and so any interference could influence the measured parameters. This is tested by performing two additional fits, independently allowing interference between each pair of adjacent states, and assigning a systematic uncertainty equal to the maximum shift in each parameter compared to the baseline result (\emph{Interference}).

Potential mismodeling of the detector resolution in simulation is tested by multiplying the Gaussian widths for all resolution functions by a scale factor of 0.9 or 1.1 and repeating the fits. This range is chosen based on detailed studies of the $\Xicp(2645)$ peak in data and simulation. Systematic uncertainties on fit parameters are then assigned as their maximum shifts  (\emph{Resolution}).
The description of the background component in the fit to the $\Delta m$ distribution is varied by using a third-order polynomial function, and associated systematic uncertainties are assigned as the shift in fit parameters from their baseline values (\emph{Background}).

\begin{table}
  \caption{Summary of uncertainties on all measured parameters (in \mev), including a breakdown of systematic uncertainties as described in the text. Entries marked with a dash are not relevant for that measurement.}
    \begin{center}\begin{tabular}{l|cc|cc|cc|cc}
    \hline
        \multirow{2}{*}{Contribution} & \multicolumn{2}{c|}{\Xicststp{2815}} & \multicolumn{2}{c|}{\Xicststp{2923}} & \multicolumn{2}{c|}{\Xicststp{2970}} & \multicolumn{2}{c}{\Xicststp{3080}} \\
        & $m$             & $\Gamma$            & $m$             & $\Gamma$            & $m$             & $\Gamma$            & $m$            & $\Gamma$            \\ \hline
    RBW          & $0.00$ & $0.02$ & $0.09$ & $1.12$ & $0.15$ & $0.55$ & $0.02$ & $0.68$ \\
    Interference & $0.00$ & $0.01$ & $0.53$ & $0.70$ & $0.44$ & $1.05$ & $1.26$ & $0.65$ \\
    Resolution   & $0.00$ & $0.12$ & $0.00$ & $0.11$ & $0.01$ & $0.04$ & $0.00$ &  $0.08$ \\
    Background   & $0.00$ & $0.01$ & $0.01$ & $0.32$ & $0.11$ & $1.51$ & $0.04$ &  $0.08$ \\
    Mom. scale   & $0.02$ &  ---   & $0.05$ &  ---   & $0.07$ &  ---   & $0.10$ &    ---    \\
    Energy loss  & $0.02$ &  ---   & $0.02$ &  ---   & $0.02$ &  ---   & $0.02$ &    ---    \\
    Sel. bias    & $0.02$ &  ---   & $0.02$ &  ---   & $0.02$ &  ---   & $0.02$ &    ---    \\ \hline
    Total syst.  &   0.03 &   0.12 &   0.54 &   1.36 &   0.48 &   1.92 &   1.27 &    0.95 \\ 
    Stat.        &   0.03 &   0.08 &   0.28 &   0.87 &   0.46 &   1.68 &   0.72 &     2.28 \\ 
    \hline
    \end{tabular}\end{center}
    \label{tab:syst}
\end{table}

The mass measurements may be affected by the LHCb momentum scale calibration (\emph{Mom. scale}) and by mismodeling of material leading to incorrect energy loss assumptions in simulation and particle reconstruction (\emph{Energy loss}). Previous studies indicate that the former effect is associated with a systematic uncertainty corresponding to 0.03\% of the Q-value for the decay~\cite{LHCb-PAPER-2013-011}. The systematic uncertainty due to imperfect modeling of the energy loss is assigned as 0.02\mev for all mass measurements, corresponding to 0.01\mev for each of the two pion tracks entering the $\Delta M$ calculation~\cite{LHCb-PAPER-2013-011}. The impact from other final-state particles cancels in the mass difference.
In addition, a potential bias in the mass measurements originating from the event selection is tested by comparing the reconstructed masses in simulation before and after the offline selection requirements are imposed. A small shift of 0.016\mev is found, consistent for all four states, which is assigned as a systematic uncertainty (\emph{Sel. bias}). No correction is applied as this shift is found to be compatible with zero.

Several cross-checks are performed to confirm the robustness of the results. The fit is repeated separately for each of the two $\Xires_c^{\pm}$ charges, for each of the two LHCb dipole magnet polarities, and for each of the three data-taking years. In all cases the results are self-consistent. Fits are also performed with additional resonances allowed to contribute, namely \Xicststp{2939}, \Xicststp{3055}, and \Xicststp{3123} states. These are added individually to the fit, with their masses and widths constrained from external measurements~\cite{PDG2024}. In all cases the results are stable compared to the baseline fit, and no evidence is found for the presence of these extra states in the data sample.

%%% Conclusion

In conclusion, this Letter presents a search for excited \Xicp states decaying to $\Xicststz{2645}\pip$, with $\decay{\Xicststz{2645}}{ \Xicp \pim}$ and $\decay{\Xicp}{ \proton\Km\pip}$. Four states are observed with high significance, including the discovery of the \Xicststp{2923} baryon, and the first observation of the \Xicststp{3080} baryon in this final state. Masses and widths are measured for all four states, and are competitive with, or more precise than the existing world averages.
The natural width of the \Xicststp{2970} state is not consistent with that of the \Xicststz{2965} baryon observed in the \Lc \Km mass spectrum~\cite{LHCb-PAPER-2020-004}, indicating that the two are different excited \Xic states. 
This continues the success of the LHCb experiment in discovering new hadronic states and precisely measuring their properties, a key input for understanding QCD in the nonperturbative realm.

\section*{Acknowledgements}
%
% These Acknowledgements valid from 3-May-2019
%
\noindent We express our gratitude to our colleagues in the CERN
accelerator departments for the excellent performance of the LHC. We
thank the technical and administrative staff at the LHCb
institutes.
We acknowledge support from CERN and from the national agencies:
ARC (Australia);
CAPES, CNPq, FAPERJ and FINEP (Brazil); 
MOST and NSFC (China); 
CNRS/IN2P3 (France); 
BMBF, DFG and MPG (Germany); 
INFN (Italy); 
NWO (Netherlands); 
MNiSW and NCN (Poland); 
MCID/IFA (Romania); 
%MSHE (Russia); 
MICIU and AEI (Spain);
SNSF and SER (Switzerland); 
NASU (Ukraine); 
STFC (United Kingdom); 
DOE NP and NSF (USA).
%%%%%%%%%%%%%%%%%%%%%%%%%%%%%%%%%%%%%%%%%%%%%
We acknowledge the computing resources that are provided by ARDC (Australia), 
CBPF (Brazil),
CERN, 
IHEP and LZU (China),
IN2P3 (France), 
KIT and DESY (Germany), 
INFN (Italy), 
SURF (Netherlands),
Polish WLCG (Poland),
IFIN-HH (Romania), 
%RRCKI and Yandex LLC (Russia), 
PIC (Spain), CSCS (Switzerland), 
and GridPP (United Kingdom).
%%%%%%%%%%%%%%%%%%%%%%%%%%%%%%%%%%%%%%%%%%
We are indebted to the communities behind the multiple open-source
software packages on which we depend.
%%%%%%%%%%%%%%%%%%%%%%%%%%%%%%%%%%%%%%%%%%
Individual groups or members have received support from
%ARC and ARDC (Australia); % moved to national 16/01
Key Research Program of Frontier Sciences of CAS, CAS PIFI, CAS CCEPP, 
Fundamental Research Funds for the Central Universities,  and Sci.\ \& Tech.\ Program of Guangzhou (China);
Minciencias (Colombia);
EPLANET, Marie Sk\l{}odowska-Curie Actions, ERC and NextGenerationEU (European Union);
A*MIDEX, ANR, IPhU and Labex P2IO, and R\'{e}gion Auvergne-Rh\^{o}ne-Alpes (France);
%RFBR, RSF and Yandex LLC (Russia);
Alexander-von-Humboldt Foundation (Germany);
ICSC (Italy); 
%GVA, XuntaGal, GENCAT, Inditex, InTalent and Prog.~Atracci\'on Talento, CM (Spain);
Severo Ochoa and Mar\'ia de Maeztu Units of Excellence, GVA, XuntaGal, GENCAT, InTalent-Inditex and Prog.~Atracci\'on Talento CM (Spain);
SRC (Sweden);
the Leverhulme Trust, the Royal Society and UKRI (United Kingdom).

  %taken from EB gitlab 19 Sep 2024
\clearpage
\section*{End Matter}

\subsection*{Additional information on multivariate selection}

Three sequential multivariate algorithms (MVA) are used to select signal candidates for this analysis. 
%Each is independently optimised for the specific purpose.
This gives better performance than using a single MVA, because the classifiers are tuned to reject the particular background contributions at each stage, and the first two benefit from being trained with real data. The classifiers are selected independently based on the best performance, and are drawn from several different machine learning libraries.

The first MVA maximises the significance of the ground state \Xicp signal, discriminating against random three-track combinations which pass the earlier selection criteria. The best performance is obtained using a feed-forward neural network from the {\sc TensorFlow} and \texttt{Keras} libraries, with a [$n$,32,16,4,1] topology, where $n=29$ is the number of input variables. A Leaky Rectified Linear Unit (\texttt{LeakyReLU}) activation function is used for the hidden layers [32,16,4] to improve the performance.
%, with a sigmoid activation function for the output layer to give a classifier response between 0 and 1. 
The 29 input variables are:
\begin{itemize}
    \item {\bf Topological variables (10):} Fit $\chi^2$ of the \Xicp decay vertex; Flight distance (FD) of the \Xicp candidate; FD/uncertainty of the \Xicp candidate; Difference in PV fit $\chi^2$ with and without the \Xicp included in the fit ($\chi^2_{\text{IP}}$); Cosine of the direction angle between the \Xicp displacement and momentum vectors; Distance-of-closest-approach (DOCA) between each pair of final-state particles; Minimum DOCA value; Impact parameter of the proton.
    \item {\bf Kinematic variables (2):} \Xicp pseudorapidity; proton transverse momentum.
    \item {\bf Particle identification variables (17):} For each final-state hadron: The output of LHCb-specific multivariate classifiers used to separate different hadron hypotheses using information from the RICH and other subdetectors, and the output of a separate multivariate algorithm to reject fake tracks from random hit combinations.
\end{itemize}
The classifier is trained with data, using the \sPlot method as described in the main text.

The second MVA maximises the significance of the $\Xic(2645)^0$ signal, discriminating against a background primarily coming from genuine \Xicp baryons combined with an unrelated pion. The best performance is obtained with a gradient boosted decision tree from the \texttt{sklearn} library, using 14 input variables:
\begin{itemize}
    \item {\bf Topological variables (6):} Reduced $\chi^2$ of the kinematic refit that constrains the pion and \Xicp to originate from a common vertex; Reduced $\chi^2$ of the $\Xic(2645)^0$ decay vertex fit; $\chi^2_{\text{IP}}$ for the $\Xic(2645)^0$ candidate, the pion from the $\Xic(2645)^0$ decay, and the \Xicp candidate.
    \item {\bf Kinematic variables (3):} $\Xic(2645)^0$ momentum; Momentum and transverse momentum of the pion from the $\Xic(2645)^0$ decay.
    \item {\bf Particle identification variables (4):} For the pion from the $\Xic(2645)^0$ decay, outputs from LHCb-specific multivariate classifiers used to differentiate between pion and other hadron hypotheses, and the output of a separate multivariate algorithm to reject fake tracks from random hit combinations.
    \item {\bf Other variables (1):} The output of the stage-1 MVA classifer.
    \end{itemize}
    This classifier is also trained using the \sPlot method with data, as described in the main text.

The third MVA is optimised using the Punzi figure-of-merit, to discriminate the \Xicpst signals from a hybrid background including both genuine and fake $\Xic(2645)^0$ candidates combined with an unrelated pion. The best performance is obtained with a gradient boosted decision tree from the \texttt{catboost} library, with 15 input variables:
\begin{itemize}
    \item {\bf Topological variables (8):} Vertex-fit $\chi^2$ of the \Xicpst candidate; $\chi^2_{\text{IP}}$ of the \Xicpst candidate; $\chi^2_{\text{IP}}$ of each of the two pions from the \Xicpst decay; FD/uncertainty of the \Xicpst candidate; FD/uncertainty of the \Xicp candidate; Reduced $\chi^2$ of the kinematic refit that constrains the \Xicp candidate and the two accompanying pions to originate from a common vertex; A derived quantity $\chi^2_{\text{IP}}(\Xicp) / [\chi^2_{\text{IP}}(\Xicp) + \chi^2_{\text{IP}}(\pip) + \chi^2_{\text{IP}}(\pim)]$.
    \item {\bf Kinematic variables (3):} Three quantities derived from the transverse momenta of the \Xicpst decay products: [$\pt(\Xicp) + \pt(\pip) + \pt(\pim)$]; min$(\pt(\pip),\pt(\pim))$; $\pt(\Xicp)/[\pt(\Xicp) + \pt(\pip) + \pt(\pim)]$.
    \item {\bf Particle identification variables (4):} For each of the two pions from the \Xicpst decay, the value of two LHCb multivariate classifiers respectively quantifying the confidence in a pion hypothesis, and in a fake-track hypothesis. 
    \end{itemize}
    The classifier is trained using simulation and control samples in data, as described in the main text.

\subsection*{Mass distributions for wrong-sign samples}

\begin{figure}
    \centering
    \includegraphics[width=0.8\linewidth]{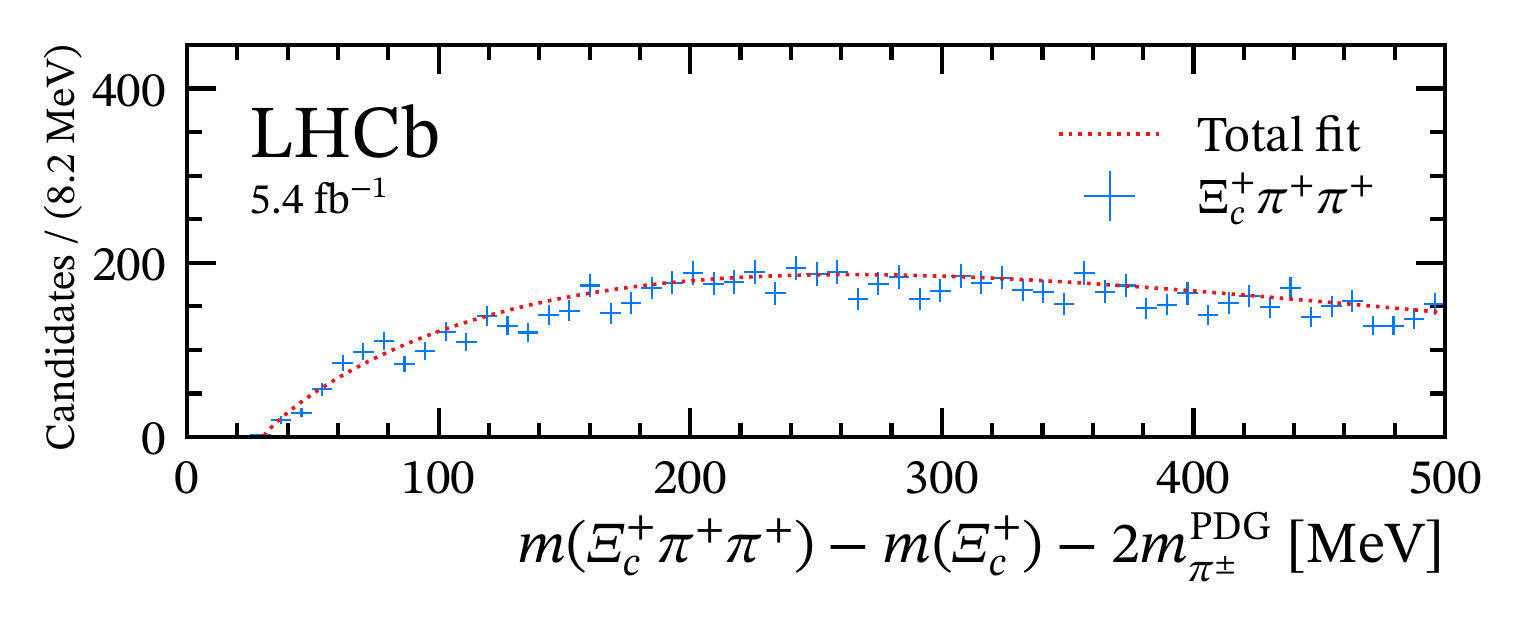}
    \caption{Distribution of $\Delta M$ for candidates from the wrong-sign $\Xicp\pip\pip$ sample, passing all selection requirements. No peaking behaviour is observed. An illustrative fit is also shown using the same function as used for the signal sample.}
    \label{fig:WS-DeltaM}
\end{figure}

To verify that the selection criteria do not cause any distortions of the $\Delta M$ distribution, the wrong-sign sample is inspected after applying all requirements. The result is shown in Fig.~\ref{fig:WS-DeltaM}. The overall impact of the selection is to suppress background fairly uniformly across the $\Delta M$ range, and there is no indication of any peaking behaviour. An illustrative fit is performed with the same background model as used for the signal sample.

\subsection*{Mass difference measurements}

In the fit to the $\Delta M$ distribution, the peak positions are free parameters, denoted $\delta M(X)$ for state $X$. The values returned by the fit are
\begin{align*}
    \delta M[\Xic(2815)^{+}] &=     \XicOneDelta,  \\ 
    \delta M[\Xic(2923)^{+}] &=     \XicTwoDelta, \\  
    \delta M[\Xic(2970)^{+}] &=     \XicThreeDelta, \\
    \delta M[\Xic(3080)^{+}] &=     \XicFourDelta,  
\end{align*}
where the first uncertainty is statistical and the second is due to experimental systematic effects. 
These measurements are converted into absolute mass measurements for the excited baryons, as reported in the paper, by adding the masses of the two charged pions and the \Xicp baryon ($2467.71 \pm 0.23 \mev$) taken from Ref.~\cite{PDG2024}.

\clearpage

%%%%%%%%%%%%%%%%%%%%%%%%%
%%%%% Bibliography %%%%%%
%%%%%%%%%%%%%%%%%%%%%%%%%

\clearpage
\addcontentsline{toc}{section}{References}
\bibliographystyle{LHCb}
\bibliography{main,standard,LHCb-PAPER,LHCb-CONF,LHCb-DP,LHCb-TDR}

%%%%%%%%%%%%%%%%%%%%%%%%%
%%%%% Author list %%%%%%%
%%%%%%%%%%%%%%%%%%%%%%%%%
 
\newpage
% LHCb collaboration author list
% Data extracted on February 21st, 2025 at 3:35pm for paper reference LHCb-PAPER-2024-055
\centerline
{\large\bf LHCb collaboration}
\begin
{flushleft}
\small
R.~Aaij$^{38}$\lhcborcid{0000-0003-0533-1952},
A.S.W.~Abdelmotteleb$^{57}$\lhcborcid{0000-0001-7905-0542},
C.~Abellan~Beteta$^{51}$\lhcborcid{0009-0009-0869-6798},
F.~Abudin{\'e}n$^{57}$\lhcborcid{0000-0002-6737-3528},
T.~Ackernley$^{61}$\lhcborcid{0000-0002-5951-3498},
A. A. ~Adefisoye$^{69}$\lhcborcid{0000-0003-2448-1550},
B.~Adeva$^{47}$\lhcborcid{0000-0001-9756-3712},
M.~Adinolfi$^{55}$\lhcborcid{0000-0002-1326-1264},
P.~Adlarson$^{83}$\lhcborcid{0000-0001-6280-3851},
C.~Agapopoulou$^{14}$\lhcborcid{0000-0002-2368-0147}
% ...

{\footnotesize \it

% ...
$^{14}$Universit{\'e} Paris-Saclay, CNRS/IN2P3, IJCLab, Orsay, France\\
% ...
$^{38}$Nikhef National Institute for Subatomic Physics, Amsterdam, Netherlands\\
% ...
$^{47}$Instituto Galego de F{\'\i}sica de Altas Enerx{\'\i}as (IGFAE), Universidade de Santiago de Compostela, Santiago de Compostela, Spain\\
% ...
$^{51}$Physik-Institut, Universit{\"a}t Z{\"u}rich, Z{\"u}rich, Switzerland\\
% ...
$^{55}$H.H. Wills Physics Laboratory, University of Bristol, Bristol, United Kingdom\\
% ...
$^{57}$Department of Physics, University of Warwick, Coventry, United Kingdom\\
% ...
$^{60}$School of Physics and Astronomy, University of Glasgow, Glasgow, United Kingdom\\
$^{61}$Oliver Lodge Laboratory, University of Liverpool, Liverpool, United Kingdom\\
% ...
$^{69}$Syracuse University, Syracuse, NY, United States\\
% ...
$^{83}$Department of Physics and Astronomy, Uppsala University, Uppsala, Sweden, associated to $^{60}$\\
% ...
}
\end{flushleft}

\end{document}